# Graphene Manipulation on 4H-SiC(0001) Using Scanning Tunneling Microscopy

Peng Xu[1], Matthew. L. Ackerman[1], Steven D. Barber[1], James K. Schoelz[1], Dejun Qi[1], Paul M. Thibado[1], Virginia D. Wheeler[2], Luke O. Nyakiti[2], Rachael L. Myers-Ward[2], Charles R. Eddy, Jr.[2], and D. Kurt Gaskill[2]

[1]*Department of Physics, University of Arkansas, Fayetteville, AR 72701, U.S.A.*

[2]*Naval Research Laboratory, Washington, D.C. 20375, U.S.A.*

Atomic-scale topography of epitaxial multilayer graphene grown on 4H-SiC(0001) was investigated using scanning tunneling microscopy (STM). Bunched nano-ridges ten times smaller than previously recorded were observed throughout the surface, the morphology of which was systematically altered using a relatively new technique called electrostatic-manipulation scanning tunneling microscopy. Transformed graphene formations sometimes spontaneously returned to their original morphology, while others permanently changed. Using an electrostatic model, we calculate that a force up to ~5 nN was exerted by the STM tip, and an energy of around 10 eV was required to alter the geometry of a ~100 × 200 nm$^2$ area.

## 1. Introduction

Epitaxial graphene grown through the thermal decomposition of SiC shows perhaps the most commercial promise for application in carbon-based electronics.[1] In the semi-insulating form, SiC would not shunt the current flow in graphene, and it is already available in the form of large-diameter wafers compatible with current industrial technology.[2-4] However, low electron mobility is a significant problem and is thought to be due to strong interactions with the substrate.[5] Using atomic-scale scanning tunneling spectroscopy (STS), the source of scattering was revealed to be inhomogeneity in the electronic structure of the graphene-SiC interface.[6]

In addition to electronic properties, interfacial forces acting on epitaxial graphene also strongly influence its topographical structure. For example, large-scale atomic force microscopy images of epitaxial graphene on Si-face SiC have revealed large ridges (1-2 nm high) running parallel to steps in the substrate.[7-9] Atomic-scale scanning tunneling microscopy (STM) images have discovered that the ridges are buckled regions of the graphene, and sometimes randomly rearrange under the influence of the STM tip during imaging.[10] In transferred graphene on a $SiO_2$ substrate, Mashoff *et al.* were able using STM to induce movement in the nano-membranes and even observe bistability.[11] In a related study, an STM technique called electrostatic-manipulation STM (EM-STM) has been used to controllably displace freestanding graphene[12] and to locally separate the top layer of graphite from the bulk.[13,14] EM-STM is similar to constant-current STS, in that the bias voltage is varied as one records the vertical displacement of the tip required to maintain a constant current. Assuming the sample is stationary, this process indirectly probes its density of states (DOS). However, the biased tip also induces an image charge in the grounded sample, resulting in an electrostatic attraction that increases with bias voltage. This force sometimes causes movement of the surface, convoluting and often eclipsing any DOS measurement. This process is illustrated in Fig. 1(a), which shows a graphene layer on a SiC substrate being locally lifted by the electrostatic attraction to the STM tip. By utilizing this force, one may not only physically manipulate a sample and examine the energetics of both its mechanical and electrostatic properties, but one can also control or alter the geometry of the interface between the thin film and the substrate.[12]

In this study, STM images reveal bunched nano-ridges ~0.1 nm high throughout the multilayer graphene grown on a 4H-SiC(0001) substrate. EM-STM measurements are performed and are found to introduce both reversible and irreversible local changes in the morphology.

## 2. Experiment

The epitaxial multilayer graphene sample used in this study was grown on 20 μm of intentionally n-doped ($1\times10^{14}$/cm$^3$) epitaxial 4H-SiC layer on the Si face of a 4º off-cut 4H-SiC substrate (Dow Corning) measuring 16 × 16 mm$^2$ and cut from a 76.2 mm diameter parent wafer. Growth was carried out in a commercially available hot-wall Aixtron VP508 chemical vapor deposition reactor. Prior to graphene growth, the substrate was etched *in situ* in a H$_2$ ambient environment for 5 min at 1520 °C. This etching produces a controlled starting surface that is dominated by SiC surface steps roughly 0.5 nm high. After the H$_2$ etching step, the ambient environment was switched to Ar with a transition period of 2 min during which pressures varied by ±50% around 100 mbar. The subsequent 120 min graphene growth process was conducted under a flowing Ar environment of 20 slm at 100 mbar, with a substrate growth temperature of 1620 °C.[15] After growth, the sample was cooled to room temperature, cut to 7 × 14 mm$^2$ in size, diamond scribed with labels on the carbon face and transported to the STM facility. Constant-current STM images and EM-STM data were obtained using an Omicron ultrahigh-vacuum (base pressure is 10$^{-10}$ mbar), STM (low-temperature model) operated at room temperature. The sample was mounted into the STM chamber where it was electrically grounded. STM tips were electrochemically etched from 0.25 mm diameter polycrystalline tungsten wire. All STM images were acquired using a positive tip bias of 0.1 V and a tunneling current setpoint of 0.05 nA.

## 3. Results and Discussion

For the simulation component, we calculated the electrostatic attractive force between the tip and sample using a highly idealized model.[16] The STM tip is replaced by a conducting sphere of radius *a* (20 nm), held at potential $V$, and placed with its surface a distance $d$ (1 nm) from an infinite grounded conducting plane representing the sample, as illustrated in Fig. 1(b). A simulation result showing the electric field lines after including three charge pairs (also illustrated) and $V = 1.0$ V is shown in Fig. 1(b). From the field lines, we estimate the affected sample radius to be about $5a$. However, the force pulling the graphene can extend far beyond this distance because remote areas with excess graphene will move.

A filled-state STM image measuring 320 nm × 320 nm is shown in Fig. 1(c). The STM data was acquired from a (0001) facet, so no SiC steps are present in this image. The topographic features are graphene nano-ridges, which are ten times lower than the higher ridges previously reported. The graphene is held flat throughout the left edge of the image, while some undulations occur in the middle, most notably creating two vertical trenches with a narrow elevated ridge between them. A line profile, drawn underneath, shows a height change of ~0.12 nm from black to white. Immediately after the scan was completed, a series of EM-STM measurements were performed in a pattern along the narrow raised strip near the middle of the image, using a setpoint current of 0.05 nA and a voltage ramp from 0.1 to 3.0 V. Then, a second STM image of the same size and same location was acquired and is shown in Fig. 1(d). Significant changes in the surface morphology occurred as a result of the EM-STM measurements, especially near the middle. The entire trench/ridge structure has essentially inverted. The middle section surprisingly maintains the same boundary design but high spots have become low, while low spots have become high. A line profile shows similar height changes to before. Additional EM-STM measurements were performed in this same area with no further changes.

The EM-STM data taken along the nano-ridge was averaged together and displayed in Fig. 1(e). It shows that the height of the STM tip continuously increased up to 1.25 nm as the bias voltage was increased to 3.0 V. The inset shows that the average tunneling current ($I$) remained roughly constant at 0.05 nA throughout the voltage ($V$) ramp. From basic tunneling theory, it is expected that the STM tip would retract to maintain a constant current as the bias voltage increases. However, this model also assumes the sample does not move. Given that the morphology of the sample has changed, modeling the height-voltage data is not possible.

Nevertheless, we can model the forces and energies involved in the EM-STM process. We find from the simple model the force to be $F = 0.5\,V^2$ with a = 20 nm and d = 1 nm, and it is plotted as an inset in Fig. 1(f). Note that the electrostatic force only weakly depends on the tip-sample separation. The force curve was used to convert the height-voltage data into force-height data as shown in Fig. 1(f). The force increases almost linearly with height to a maximum of ~5 nN at a tip height of 1.25 nm. The area under the force-height curve reveals that an energy cost of around 13 eV was expended by the STM to alter the shape of about $100 \times 200$ nm$^2$ sized graphene-substrate interface [size estimated from the central region of Fig. 1(c)]. Calculations done previously show the force and energy required to separate a single unit cell of graphene from graphite is ~0.1 nN and ~50 meV, respectively.[13] This is consistent with our estimates given the total area affected, the starting roughness (i.e., contact area), and that the graphene is predominately sliding across the surface.

EM-STM measurements were carried out on another part of the sample showing nano-ridges in a chronological sequence as illustrated in Fig. 2. Each filled-state STM image in Figs. 2(a)-2(e) shows the same $320 \times 320$ nm$^2$ area, coupled with a line profile taken along the marked location on the image. Notice the three diagonally-running parallel ridges aligned from the bottom left corner to the top right corner of the starting image in Fig. 2(a). On average, they are ~0.1 nm high and 30 nm wide. The kinks provide a marker for future reference. The subsequent scan [shown in Fig. 2(b)] was paused several lines in (notice the point of significant contrast change near the bottom), and an EM-STM measurement was taken near the center of the image directly on top of the central diagonal ridge. The scan was then immediately resumed where it had left off. Afterward the trenches are significantly flattened in the STM image. The height changes across the surface are significantly reduced compared to before the EM-STM (see line profiles below). The very next image taken is shown in Fig. 2(c) and, surprisingly, it is almost identical to the original image. The line profile shown below also confirms that the height changes are back to normal. The image is slightly noisier than the original, but we believe this is a true instability in the surface being captured in the STM data. To demonstrate reproducibility, we repeated the previous procedure as shown in Figs. 2(d) and 2(e). Again, the graphene layer has been flattened and recovered. The main differences are an alteration in the upper left corner and that the ridges are not quite as high. The two EM-STM measurements taken during the chronological sequence are shown in Fig. 2(f), offset by 0.2 nm for clarity. The shapes of the two curves are similar and show a nearly linear dependence with tip bias. There is an interesting reproduced abrupt jump above two volts, indicating that the graphene lurched suddenly at this voltage level.

The presence of the nano-ridges and their physical separation from the SiC substrate is not unexpected given the presence of the larger ridges previously reported.[17] We are certain the formation mechanism is the same. The appearance of these ridges has been attributed to the stress in the surface layer caused by graphene expanding and SiC contracting as the system cools from its high growth temperature.[18-20] Note, we have confirmed that the ridges are topographic features and not electronic features using alternate bias STM imaging.[21,22] After the cool down process, there are two dominant forces acting on the graphene. The first is the attractive van der Waals interaction between the graphene and the under layer, which would try to maximize the area of the graphene lying flat on the atomically smooth substrate. The second is the compressive stress induced in the graphene because it expands as the SiC substrate contracts during cooling. Naturally, at edges and defect sites, the graphene is more strongly bonded causing other nearby locations to bow out into the vacuum away from the atomically smooth substrate. For a particular cool down, there could be numerous possible morphologies that balance the two competing forces. However, some morphologies are likely to be kinetically limited geometries. Only after the EM-STM measurement could this area change to a lower energy morphology. For the region of the sample shown in Fig. 2, this was not the case. Here, the minimum energy configuration was achieved at the beginning, so the surface consistently regained its initial geometry following perturbation.

Another special feature of graphene is that the ridges will create local pseudo-magnetic fields, which will alter the transport properties.[23] These fields are strongest in regions of highest curvature. It is more difficult experimentally to find or engineer suitable test structures,[24] but the structures observed here might be explored as possible candidates.

Ultimately, when the graphene is lifted off the substrate by the electrostatic force during an EM-STM measurement, van der Waals forces at the interface must be overcome, and the required energy is equal to the work done. Thus, this technique can not only allow for local manipulation of the graphene surface, but it can also give valuable information about bonding at the graphene-substrate interface, which plays such a central role in charge puddling, scattering, and electron mobility.[25]

## 4. Summary

In summary, multilayer epitaxial graphene grown on 4H-SiC(0001) was studied using STM. Nano-ridges around 0.1 nm high were observed and found to be maneuverable using EM-STM measurements carried out on top of the nano-ridges. In some areas of the sample, permanent changes could be induced through the interaction between the STM tip and the sample. An electrostatic model was used to quantify this interaction, showing that an attractive force up to ~5 nN is distributed over a ~100 nm × 200 nm area and that energies of about 10 eV are expended in the EM-STM manipulations.

**Acknowledgements**

P.X. and P.T. gratefully acknowledge the financial support of ONR under grant N00014-10-1-0181 and NSF under grant DMR-0855358. Work at the U.S. Naval Research Laboratory is supported by the Office of Naval Research. LON gratefully acknowledges postdoctoral fellowship support through the ASEE.

Fig. 1: (Color online) (a) Schematic of STM tip lifting graphene off SiC. (b) Calculated electric field lines shown leaving a biased conducting sphere (STM tip) and ending at a grounded plane (graphene sample). (c,d) 320 × 320 nm² filled-state STM images showing before and after EM-STM, respectively. Associated line profiles are displayed below the images. (e) Average tip height vs. bias during the EM-STM measurement. Inset: average tunneling current vs. bias. (f) The electrostatic attractive force between the tip and sample vs. tip height. Inset: calculated force vs. tip bias curve used to transform the data.

Fig. 2: (Color online) A chronological series of filled-state STM images 320 × 320 nm² (0.1 V, 0.05 nA). A line profile is provided below each image from the marked location. (a) Initial image of the surface. (b) Second image paused several lines in and an EM-STM measurement was taken. The rest of the image was then collected and reveals the morphological changes. (c) Third image taken immediately after the previous. (d) Fourth image was acquired just after a second EM-STM measurement was taken. (e) Final fifth image of the sample, taken directly after the previous. (f) Two EM-STM data sets showing tip height vs. tip bias. Inset: Atomic-scale (6 × 6 nm²), filled-state (0.1 V, 0.05 nA) STM image showing the honeycomb structure of the graphene taken from the area near the nano-ridges.

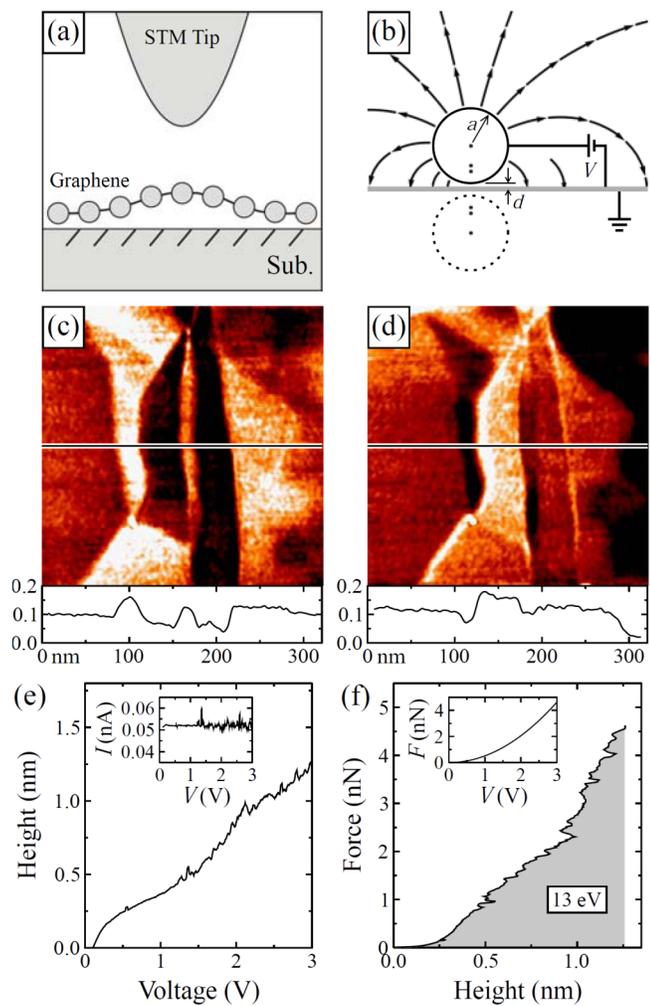

Fig. 1

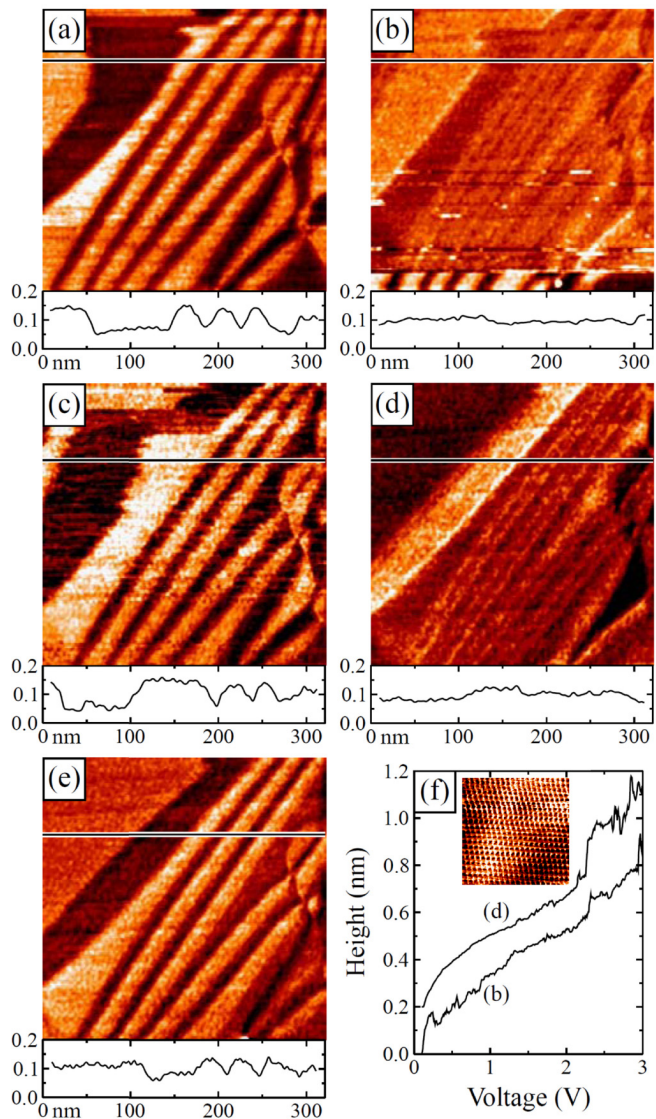

Fig. 2